\documentclass[11pt,a4paper]{article}

\usepackage[margin=1in]{geometry}
\usepackage[T1]{fontenc}
\usepackage[utf8]{inputenc}
\usepackage{lmodern}
\usepackage{microtype}
\usepackage{booktabs}
\usepackage{array}
\usepackage{graphicx}
\usepackage{float}
\usepackage{amsmath}
\usepackage{enumitem}
\usepackage{longtable}
\usepackage{pdflscape}
\usepackage{cite}
\usepackage{xurl}
\usepackage[hidelinks,breaklinks=true]{hyperref}
\graphicspath{{figures/}}

\title{Hidden urban exposure to energy-price shocks: spreading burdens and shifting hotspots}
\author{%
\normalsize
Yunfei Bai\textsuperscript{1},
Aaron Tesfa Tsion\textsuperscript{1},
Ra\'ul Rosales\textsuperscript{1},
Barbara Shollock\textsuperscript{1},
Wei He\textsuperscript{1}\thanks{Corresponding author: \href{mailto:wei.4.he@kcl.ac.uk}{wei.4.he@kcl.ac.uk}}\\[0.5em]
\normalsize\textsuperscript{1}Department of Engineering, King's College London, London, UK%
}
\date{}

\newcommand{\CurrentHighShare}{70.5}

\newcommand{\CurrentMeanProbability}{22.5}
\newcommand{\CurrentGasHighShare}{82.3}
\newcommand{\CurrentElectricHighShare}{76.9}
\newcommand{\CurrentCommonHighShare}{87.5}

\newcommand{\CurrentHiddenCount}{182}
\newcommand{\CurrentHiddenJaccard}{0.896}

\newcommand{\ElecRetained}{134}
\newcommand{\ElecEntering}{83}
\newcommand{\ElecExiting}{48}
\newcommand{\ElecHidden}{217}
\newcommand{\GasRetained}{171}
\newcommand{\GasEntering}{10}
\newcommand{\GasExiting}{11}

\newcommand{\ElecEnteringBirmingham}{30}
\newcommand{\EnterElectricShare}{55.1}
\newcommand{\RetainElectricShare}{17.7}
\newcommand{\ExitElectricShare}{5.1}

\newcommand{\EnterFlatShare}{71.4}

\newcommand{\ExitFlatShare}{24.1}
\newcommand{\EnterNewShare}{45.5}

\newcommand{\EnterEpcEG}{10.6}

\newcommand{\ExitOwnerShare}{60.7}

\newcommand{\EnterLowerIncome}{8,887}
\newcommand{\RetainLowerIncome}{8,813}

\newcommand{\EnterGasIncrement}{463}
\newcommand{\RetainGasIncrement}{1,144}

\newcommand{\EnterElecIncrement}{260}

\newcommand{\ExitMeanPninetyState}{10.3}

\newcommand{\ExitMinusRetainLowerIncome}{2,858}
\newcommand{\ExitMinusRetainLowerIncomeLow}{1,824}
\newcommand{\ExitMinusRetainLowerIncomeHigh}{3,902}

\begin{document}
\maketitle

\begin{abstract}
Cities need to anticipate where energy-price shocks could intensify hardship beyond established hotspots. How income distributions and fuel-price balances jointly shape this hidden exposure remains unclear. Here we address this gap by stress-testing nearly one million properties in six English local authorities, combining required-energy costs for existing housing with modelled local income distributions. Under gas-led crisis prices, 41\% of neighbourhoods in the highest local exposure quarter lie outside the highest official fuel-poverty quarter. Electricity-led stress changes this hidden geography: when electricity unit prices alone triple from 2021, comparable to the unmitigated late-2022 cap increase, 38\% of the resulting hidden-exposure group are absent from the crisis-based group. Entering neighbourhoods have newer, more efficient housing, more flats and greater electricity dependence than retained areas. Crisis preparedness should extend household protection beyond current hardship hotspots, using fuel-specific stress tests to anticipate support needs as fuel prices and heating systems change.
\end{abstract}


\section{Introduction}

Energy-price shocks can intensify household hardship and extend affordability pressures beyond established hotspots. Modelling of the 2022 crisis across 116 countries estimated 63--113\% increases in direct and indirect household energy costs \cite{GuanEtAl2023GlobalEnergyPriceCrisis}. In England, 2.47 million households were estimated to be in fuel poverty in 2024 \cite{DESNZ2026AnnualFuelPoverty}, and the typical dual-fuel Direct Debit price cap rose by 13\% for July--September 2026 \cite{Ofgem2026JulyPriceCap}. Unequal housing, incomes and heating fuels translate these price changes into different local burdens \cite{SovacoolEtAl2023UKEnergyTransportPoverty,FetzerGazzeBishop2024EnergyPriceShocks}.

Fuel-poverty assessments guide responses to energy-related hardship \cite{DESNZ2026AnnualFuelPoverty,PittalisEtAl2026UrbanEnergyPoverty}. England's official Low Income Low Energy Efficiency (LILEE) indicator requires both low dwelling efficiency and low income after housing and modelled energy costs \cite{DESNZ2026FuelPovertyMethodology}. Published neighbourhood rates describe hardship under reference conditions, rather than responses to alternative fuel prices \cite{DESNZ2026SubregionalMethodology}. Energy-vulnerability research distinguishes present deprivation from susceptibility to future hardship \cite{BouzarovskiPetrova2015DomesticEnergyDeprivation,MiddlemissGillard2015EnergyVulnerability,BurlinsonDavillasGiuliettiPrice2024PriceResilience}. Which neighbourhoods outside established hotspots face high exposure despite apparently favourable mean incomes or housing efficiency remains unclear.

Price-shock studies establish unequal household cost increases \cite{GuanEtAl2023GlobalEnergyPriceCrisis}, expansion beyond established hotspots \cite{MulderDallaLongaStraver2023SpatialEnergyPoverty} and changes in poverty depth and inequality \cite{CroonEtAl2023EnergyPovertyGapIndices}. Property-level modelling in England and Wales links housing and tariffs to unequal bill increases, using area-average income and official fuel-poverty rates for socioeconomic comparisons \cite{FetzerGazzeBishop2024EnergyPriceShocks}. What remains unresolved is how within-neighbourhood income variation and the gas--electricity price balance jointly affect the reach of pressure and which housing configurations warrant attention. The 2021--2022 crisis was gas-led, and gas costs influence British wholesale electricity prices \cite{GuoCastagnetoGissey2021CostPassThrough}, but retail prices need not change proportionally. Heating is also electrifying: in 2023--24, 86\% of English households heated with gas, 8\% with electricity and about 1\% with heat pumps \cite{MHCLG2025EHSLowCarbonTechnologies}, and policy aims to raise the electric share \cite{DESNZ2026BoilerUpgradeScheme,DESNZ2026WarmHomesPlan,UKGov2025CleanHeatMarketMechanismRegulations}. As heating electrifies, exposure shifts from gas to electricity prices, making electricity-led stress a preparedness condition in its own right. Preparedness based on existing hotspots or a single price pattern can therefore overlook exposed neighbourhoods or retain outdated priorities.

Our study addresses this gap by combining required-energy costs with local income distributions and independent gas and electricity price stresses. We analyse 962,869 gas/electricity properties across Bristol, Manchester, Liverpool, Birmingham, Solihull and Westminster, selected for contrasting housing, heating, tenure and income structures. Exposure denotes the modelled cost of meeting standardised energy needs relative to income after housing costs. Pairing each dwelling's Energy Performance Certificate (EPC)-based costs with its fitted neighbourhood income distribution tests affordability across income levels that an area mean can conceal.

We hold housing and calibrated incomes fixed to isolate price effects, first testing whether high burdens spread while property rankings persist. Fuel Price-at-Risk (FPaR) then summarises the upper end of burden across local incomes and three equally weighted crisis price conditions. We define hidden exposure as the highest local exposure quarter outside the highest official fuel-poverty quarter, using official 2024 rates. Finally, we test whether that geography persists under July--September 2026 tariffs and electricity-led stress. Together, these comparisons identify where support may be needed beyond established hardship hotspots and which housing profiles warrant renewed assessment as the fuel-price balance changes and heating electrifies.

\section{Results}

\subsection{High-burden reach depends on the fuel-price balance}

We first assessed how widely high burden spreads across price conditions, before asking whether official hotspots capture the most exposed neighbourhoods. For each of the 962,869 gas/electricity properties, burden is structural required-energy cost divided by income after housing costs (Methods). We evaluate this at the fitted neighbourhood income tenth percentile, below which 10\% of local incomes fall. Because burden rises as income falls, this gives the 90th percentile (P90) of modelled burden. These assessments pair properties with local income distributions; actual occupants' incomes are not observed. We classify P90 values of at least 10\% as high burden.

At April--September 2021 prices (S0), 18.1\% of properties met this high-burden threshold. With fixed daily charges (standing charges) at S0, tripling gas unit prices alone placed 73.4\% of properties at high burden, versus 36.9\% for electricity alone.

\begin{figure}[htbp]
\centering
\includegraphics[width=\linewidth]{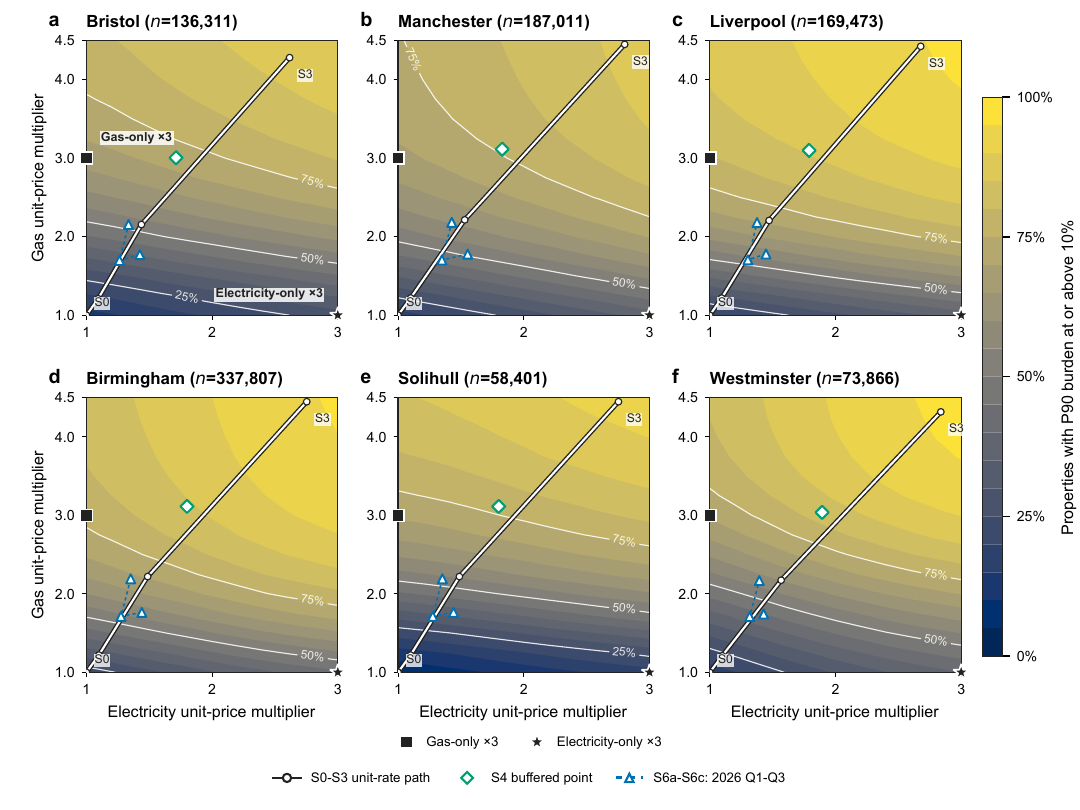}
\caption{The reach of high burden depends on gas and electricity prices. \textbf{a--f}, Share of core properties with P90 burden at or above 10\% as gas and electricity unit prices vary relative to the matching S0 prices, with zero annual standing-charge increment relative to S0. White contours mark 25\%, 50\% and 75\%. Filled squares and stars mark the gas-only and electricity-only threefold unit-price stresses, respectively, with the other fuel's unit price and standing charges held at S0. The solid path locates S0--S3 actual-region Direct Debit unit-rate ratios for each study area; the open diamond locates the S4 buffered unit-rate structure and open triangles S6a--S6c (January--September 2026). Markers locate unit-rate combinations; complete tariff outcomes appear in Fig.~\ref{fig:dynamic-stress-response}a. Required energy needs and fitted income distributions remain fixed across the specified price conditions (Methods).}
\label{fig:parametric-price-response}
\end{figure}

To separate the two fuel-price effects, these comparisons varied gas and electricity unit prices independently (Fig.~\ref{fig:parametric-price-response}). Electricity rose approximately 2.7-fold from S0 under the underlying crisis cap: the threefold electricity stress therefore has a historically grounded magnitude. Holding gas prices and standing charges at S0 isolates the effect of the electricity-price increase. Gas-only threefold prices produced high-burden shares from 67.2\% in Bristol to 79.5\% in Liverpool. The electricity-only range was wider, from 17.5\% in Solihull to 43.4\% in Birmingham.

The price increases needed to reach high burden also differed across neighbourhoods, measured as Lower layer Super Output Areas (LSOAs). Median electricity-only multipliers required for mean LSOA P90 burden to reach 10\% ranged from 2.70 in Westminster to 6.16 in Solihull. Corresponding gas-only thresholds were lower in every city (Supplementary Fig.~S8). These thresholds depend on baseline burden as well as fuel shares. This explains why Westminster was more electricity-sensitive than Manchester despite a smaller share of properties with electricity as their main fuel (Table~\ref{tab:central-area-results}).

\subsection{Gas-led stress expands burden while rankings largely persist}

\begin{figure}[htbp]
\centering
\includegraphics[width=\linewidth]{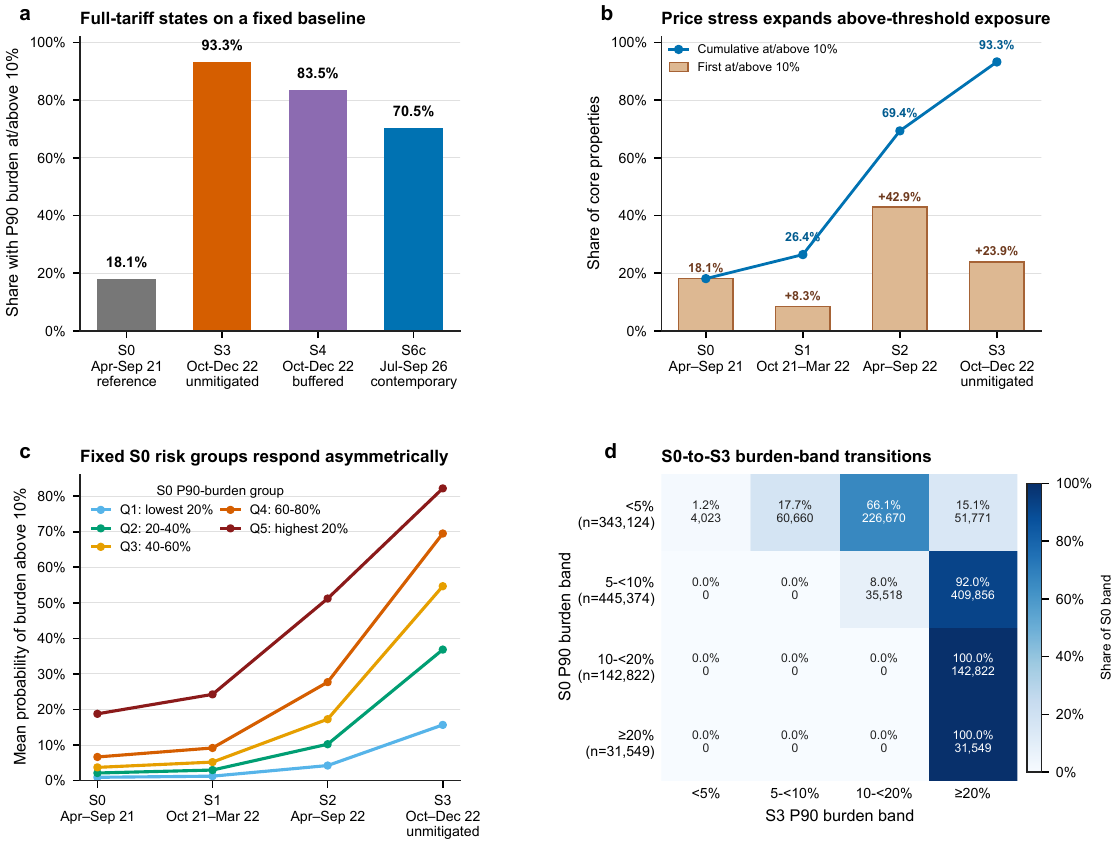}
\caption{Gas-led crisis prices spread high burden across the housing stock. \textbf{a}, Share of core properties with P90 burden at or above 10\% under four complete tariff states. S0 denotes April--September 2021 and S3 the unmitigated October--December 2022 underlying cap. S4 combines constructed Energy Price Guarantee unit rates with S3 standing charges; S6c denotes July--September 2026. Each bar applies the state's unit rates and standing-charge changes to the common baseline. \textbf{b}, Share first reaching 10\% P90 burden in each ordered S0--S3 state (bars), and cumulative share (line). \textbf{c}, Mean probability of burden above 10\% across five groups fixed from S0 burden, from Q1 (lowest fifth) to Q5 (highest fifth). \textbf{d}, Movement between S0 and S3 burden bands. Panels \textbf{a,b,d} classify properties by P90 burden; panel \textbf{c} integrates over fitted income distributions. Housing, required energy needs and fitted income distributions are held fixed throughout.}
\label{fig:dynamic-stress-response}
\end{figure}

To assess how these fuel-price effects combine, we next applied historical and 2026 tariff structures, including each period's standing charges (Fig.~\ref{fig:dynamic-stress-response}a). Compared with the unmitigated October--December 2022 underlying cap (S3), July--September 2026 tariffs combined lower gas and electricity unit prices with higher GB-average standing charges \cite{Ofgem2026DefaultTariffCapLevels,Ofgem2026JulyPriceCap}. On the common structural baseline, July--September 2026 regional Direct Debit tariffs (S6c) gave a high-burden share of \CurrentHighShare\%, compared with 18.1\% under S0. The unmitigated October--December 2022 underlying cap (S3) gave 93.3\%; applying constructed Energy Price Guarantee unit rates with S3 standing charges (S4) gave 83.5\%. The increase spanned all six study areas despite their contrasting housing and income structures (Table~\ref{tab:central-area-results}). From S6c, a further 50\% rise in unit prices with standing charges fixed raised the high-burden share to \CurrentGasHighShare\% for gas alone, \CurrentElectricHighShare\% for electricity alone and \CurrentCommonHighShare\% for both fuels (Supplementary Fig.~S11 and Table~S37).

Along the ordered crisis reference path, the largest expansion occurred at the intermediate S2 state (Fig.~\ref{fig:dynamic-stress-response}b). High-burden reach rose to 26.4\% under S1 (October 2021--March 2022) and 69.4\% under S2 (April--September 2022), before reaching 93.3\% under S3. S2 alone accounted for 42.9 percentage points of newly classified properties. By S3, 91.8\% of properties below the threshold at S0 had reached high burden.

Although almost all properties reached the 10\% threshold under S3, higher thresholds revealed differences in severity (Fig.~\ref{fig:dynamic-stress-response}c,d). The share with P90 burden of at least 20\% rose from 3.3\% under S0 to 66.1\% under S3. Mean modelled probability of burden above 10\% rose from 6.4\% to 51.8\%; under S6c, it was \CurrentMeanProbability\%. P90 classification tests exposure at the lower-income end, whereas mean probability accounts for all income levels in the fitted local distributions. Under S3, mean probability ranged from 15.7\% in the lowest baseline-burden fifth to 82.2\% in the highest fifth (Fig.~\ref{fig:dynamic-stress-response}c).

Property burden ranks nevertheless remained closely aligned between S0 and S3 (Spearman's $\rho=0.960$). Of properties in the highest baseline fifth, 90.5\% remained there under S3. Rank persistence partly reflects the model structure: with no standing-charge increment, common proportional gas and electricity unit-price increases preserve ranks exactly (Supplementary Information, section 6). Reordering along the crisis path arises from non-proportional fuel-price changes and differences in standing-charge increments.

\begin{table}[htbp]
\centering\footnotesize\setlength{\tabcolsep}{3pt}
\caption{Study-area context, stress response and local divergence. S0 is the April--September 2021 baseline, S3 the unmitigated October--December 2022 underlying-cap counterfactual and S4 the buffered comparison that applies Energy Price Guarantee unit rates with S3 standing charges.}
\label{tab:central-area-results}
\textit{a, Housing, tenure, fuel and income context of the core gas/electricity sample.}\par\smallskip
\begin{tabular}{@{}lrrrrrrrrr@{}}
\toprule
Area & Flats & Private & Social & EPC & Mains & Floor & Income & Income & Official fuel \\
 & (\%) & rent (\%) & rent (\%) & E--G (\%) & gas (\%) & area (m$^2$) & mean (\pounds k) & P10 (\pounds k) & poverty (\%) \\
\midrule
Bristol & 41.4 & 28.8 & 13.3 & 12.6 & 81.2 & 83 & 34.2 & 14.0 & 10.6 \\
Manchester & 41.9 & 28.6 & 25.9 & 9.0 & 72.8 & 78 & 27.2 & 10.1 & 12.0 \\
Liverpool & 32.3 & 23.9 & 26.5 & 10.7 & 84.2 & 81 & 26.3 & 10.8 & 11.9 \\
Birmingham & 33.3 & 20.3 & 26.5 & 15.6 & 80.8 & 82 & 26.8 & 10.3 & 13.8 \\
Solihull & 23.2 & 12.6 & 14.1 & 11.6 & 89.9 & 95 & 34.8 & 14.6 & 8.9 \\
Westminster & 83.4 & 41.4 & 15.9 & 9.9 & 77.8 & 84 & 42.3 & 10.1 & 10.2 \\
\midrule
Six-area core & 39.2 & 24.9 & 22.9 & 12.3 & 80.2 & 82 & 29.2 & 11.2 & 12.0 \\
\bottomrule
\end{tabular}
\par\medskip
\textit{b, P90 burden response and local divergence.}\par\smallskip
\begin{tabular}{@{}lrrrrrrrrr@{}}
\toprule
Area & Core & \multicolumn{3}{c}{\shortstack{Mean P90\\burden (\%)}} & \multicolumn{3}{c}{Share $\geq10\%$ (\%)} & High-exposure & Hidden LSOAs \\
\cmidrule(lr){3-5}\cmidrule(lr){6-8}
 & properties & S0 & S3 & S4 & S0 & S3 & S4 & LSOAs & $n$ (\% of high) \\
\midrule
Bristol & 136,311 & 6.8 & 26.9 & 18.9 & 12.9 & 89.3 & 76.3 & 67 & 38 (56.7) \\
Manchester & 187,011 & 7.2 & 29.4 & 20.5 & 17.5 & 87.6 & 77.8 & 74 & 33 (44.6) \\
Liverpool & 169,473 & 7.7 & 32.1 & 22.6 & 19.2 & 96.1 & 88.7 & 76 & 24 (31.6) \\
Birmingham & 337,807 & 8.2 & 34.1 & 24.0 & 22.4 & 95.9 & 87.4 & 164 & 39 (23.8) \\
Solihull & 58,401 & 5.6 & 23.8 & 16.7 & 6.1 & 93.7 & 79.9 & 34 & 26 (76.5) \\
Westminster & 73,866 & 7.3 & 28.9 & 20.3 & 16.7 & 96.4 & 84.4 & 26 & 22 (84.6) \\
\midrule
Six-area core & 962,869 & 7.5 & 30.8 & 21.6 & 18.1 & 93.3 & 83.5 & 441 & 182 (41.3) \\
\bottomrule
\end{tabular}
\par\medskip
\textit{c, Small areas carrying half of modelled high-burden risk.}\par\smallskip
\begin{tabular}{@{}lrrrr@{}}
\toprule
Area & LSOAs & S0 & S3 & S4 \\
\midrule
Bristol & 268 & 49 & 129 & 109 \\
Manchester & 295 & 75 & 123 & 110 \\
Liverpool & 302 & 70 & 137 & 121 \\
Birmingham & 659 & 172 & 299 & 259 \\
Solihull & 134 & 44 & 67 & 62 \\
Westminster & 123 & 36 & 61 & 53 \\
\midrule
Six-area core & 1,781 & 446 & 816 & 714 \\
\bottomrule
\end{tabular}
\medskip\begin{minipage}{0.98\linewidth}\footnotesize\textit{Notes:} Panel a shares are property-weighted within each area's core gas/electricity sample. Tenure proxies use the EPC tenure field, with owner-occupied and unrecorded tenure completing each row; the electricity carrier share is the complement of the mains-gas share. Floor area is the core-property mean. Income mean and P10 are medians across LSOAs of fitted income distributions calibrated at Middle layer Super Output Area (MSOA) level, in GBP thousands. Official fuel poverty is the household-weighted aggregate of the official 2024 LSOA estimates. Panel b means and high-burden shares refer to property P90 burden and are weighted over core-eligible properties; high-exposure LSOAs lie at or above the local P75 of mean property reference FPaR among the 1,754 assessable LSOAs, and hidden LSOAs are the subset below the local fuel-poverty P75. Panel c counts neighbourhoods representing at least half of each authority's modelled high-burden risk, ordered by mean high-burden probability. Every area enters at the 100\% target.\end{minipage}
\end{table}

Despite persistent property rankings, high-burden probability became less spatially concentrated (Table~\ref{tab:central-area-results}c). Within each authority, we ordered neighbourhoods by mean probability and counted those representing at least half of its total modelled high-burden risk. Total risk here is the sum of modelled property-level probabilities (Methods). The six-authority count rose from 446 under S0 to 816 under S3; the S4 buffered comparison required 714 areas. This wider geographical reach also appears in the probability-based measure, which uses the full fitted income distributions (Supplementary Tables~S17 and S18).

\clearpage
\subsection{Prospective exposure extends beyond current fuel-poverty geography}

This wider reach raises the question of whether official fuel-poverty hotspots also identify the most exposed neighbourhoods. For this comparison, we used reference FPaR to summarise exposure across the crisis price conditions: the 90th percentile of burden jointly over fitted local incomes and three equally weighted stress states, S1--S3 (Methods). Neighbourhood exposure is the mean of property reference FPaR; the full distributions are shown in Supplementary Fig.~S4a.

The comparison covered 1,754 assessable LSOAs, each with at least 50\% of the properties in its coverage sample represented in the core gas/electricity model. We measured the exposure--visibility gap as the within-authority standardised difference between log-transformed prospective exposure and official fuel-poverty rates (Methods). Positive values indicate greater prospective exposure relative to the area's official fuel-poverty rate.

\begin{figure}[!htb]
\centering
\includegraphics[width=\linewidth]{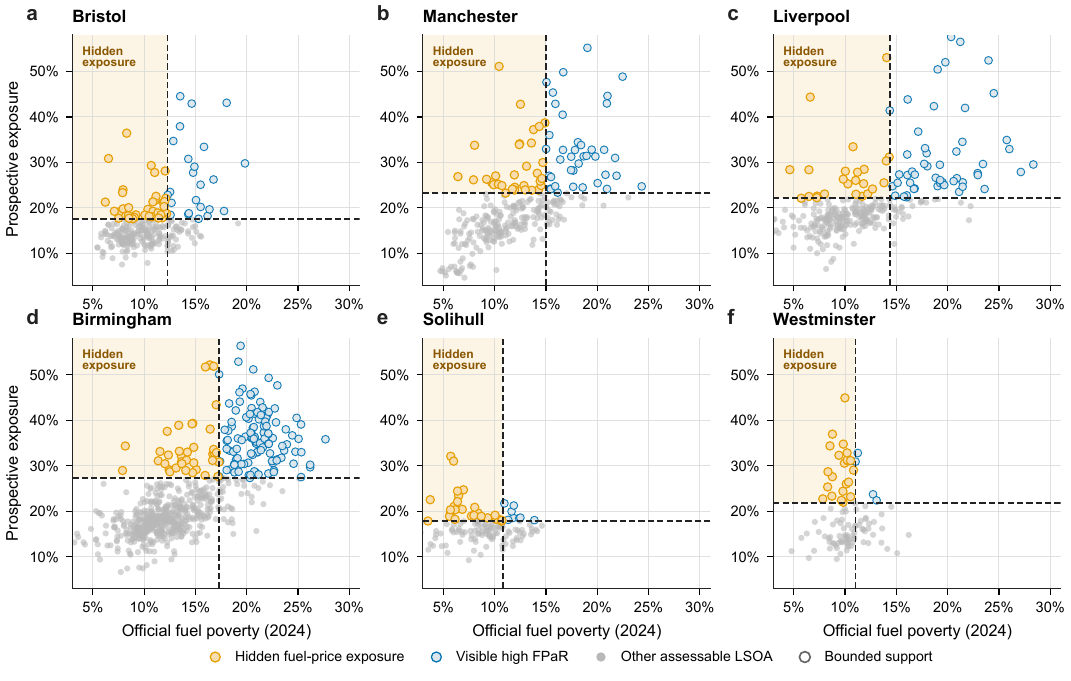}
\caption{High-exposure neighbourhoods extend beyond official fuel-poverty hotspots. Prospective exposure is mean property FPaR over the equal-weight S1--S3 reference ensemble. \textbf{a--f}, Relationship between official 2024 fuel poverty and prospective exposure among assessable small areas. Dashed lines mark the highest local quarter of each measure. Orange identifies hidden exposure, blue areas ranked highly by both measures, and grey other assessable areas. Open rings mark bounded support, with 50\% to below 80\% of coverage properties in the core model; high support is at least 80\%.}
\label{fig:hidden-exposure-divergence}
\end{figure}

The two measures placed 259 LSOAs in both local upper quarters (Fig.~\ref{fig:hidden-exposure-divergence}a--f). The remaining 182 of the 441 high-exposure areas (41.3\%) formed the reference hidden-exposure group, lying outside their authority's official fuel-poverty top quarter. This identifies a difference in local rankings, not an absence of recorded fuel poverty. This share ranged from 23.8\% in Birmingham to 84.6\% in Westminster (Table~\ref{tab:central-area-results}). Agreement exceeded chance overall (pooled Cohen's $\kappa=0.45$, using within-authority expected agreement). The matching upper-quarter cut-offs also produced 182 mirror areas with high official fuel poverty but prospective exposure below the local top quarter.

\begin{samepage}
Exposure also expanded across neighbourhoods with low official fuel poverty. Among properties in the lowest local fuel-poverty quarter, the high-burden share rose from 7.3\% under S0 to 84.9\% under S3 (Supplementary Information).

\end{samepage}

Within this broader expansion, 48.4\% of reference hidden areas already had mean P90 burden of at least 10\% at S0. The corresponding shares were 73.0\% in the visible group (high on both measures) and 0.4\% in the comparison group below both local upper-quarter cut-offs. All hidden and visible areas reached this level by S2 (Supplementary Table~S26). Hidden exposure therefore includes intensification of existing modelled burden as well as threshold crossing.

\begin{figure}[htbp]
\centering
\includegraphics[width=\linewidth]{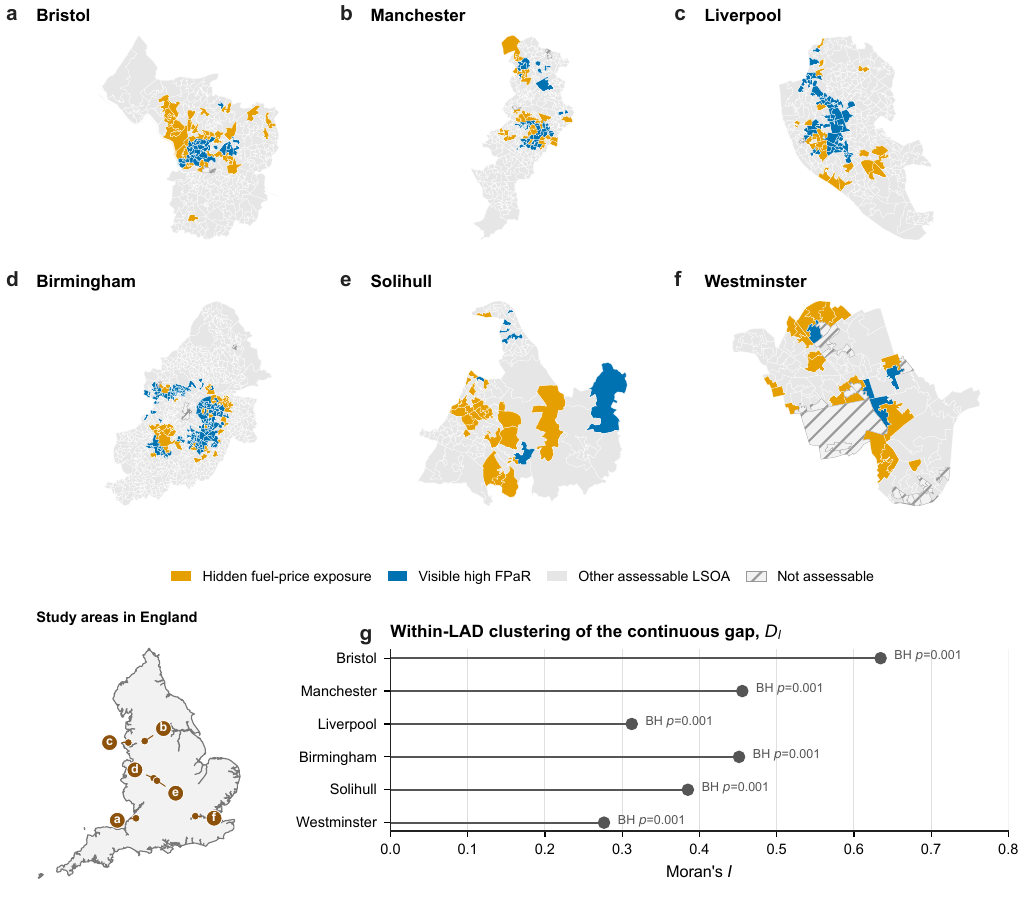}
\caption{Spatial distribution of hidden and visible fuel-price exposure. \textbf{a--f}, Fixed S1--S3 reference groups: hidden fuel-price exposure (orange; $n=182$), visible high FPaR (blue; $n=259$) and other assessable areas (grey; $n=1{,}313$). Hatched areas have less than 50\% core support and are not assessable ($n=27$). Grey includes the comparison group of 1,131 areas below both local upper-quarter cut-offs, used in Figs~\ref{fig:associative-structure} and \ref{fig:hidden-membership-transition}, and 182 mirror areas in the highest official fuel-poverty quarter. The locator marks the six study areas in England; letters correspond to panels a--f. \textbf{g}, Moran's $I$ for the continuous exposure--visibility gap $D_l$ mapped in Supplementary Fig.~S10, treating areas that share an edge or corner within each local authority as neighbours. Labels report two-sided permutation $p$ values from 999 permutations after Benjamini--Hochberg adjustment across the six LADs.}
\label{fig:spatial-misalignment}
\end{figure}

The maps locate the reference hidden-exposure group and visible high-exposure areas within each study area (Fig.~\ref{fig:spatial-misalignment}a--f). The continuous exposure--visibility gap was spatially clustered in every area (Fig.~\ref{fig:spatial-misalignment}g). Moran's $I$ ranged from 0.276 in Westminster to 0.635 in Bristol, with all six permutation tests significant after Benjamini--Hochberg adjustment ($p_{\mathrm{adj}}=0.001$). Supplementary Fig.~S10 maps the direction and magnitude of this continuous gap.

To test whether this geography depended on the reference design, we reweighted the three crisis states. Neighbourhood exposure ranks remained closely aligned with the equal-weight specification (Spearman's $\rho=0.984$--1.000). Across nine alternative weightings, the hidden-exposure group contained 179--183 areas. Jaccard overlap, the shared fraction of two lists' combined membership, ranged from 0.843 to 1.000 (Supplementary Fig.~S8d). Exact membership therefore varied more than the similar counts alone suggest.

\subsection{Hidden exposure reflects different income and housing configurations}

\begin{figure}[htbp]
\centering
\includegraphics[width=\linewidth]{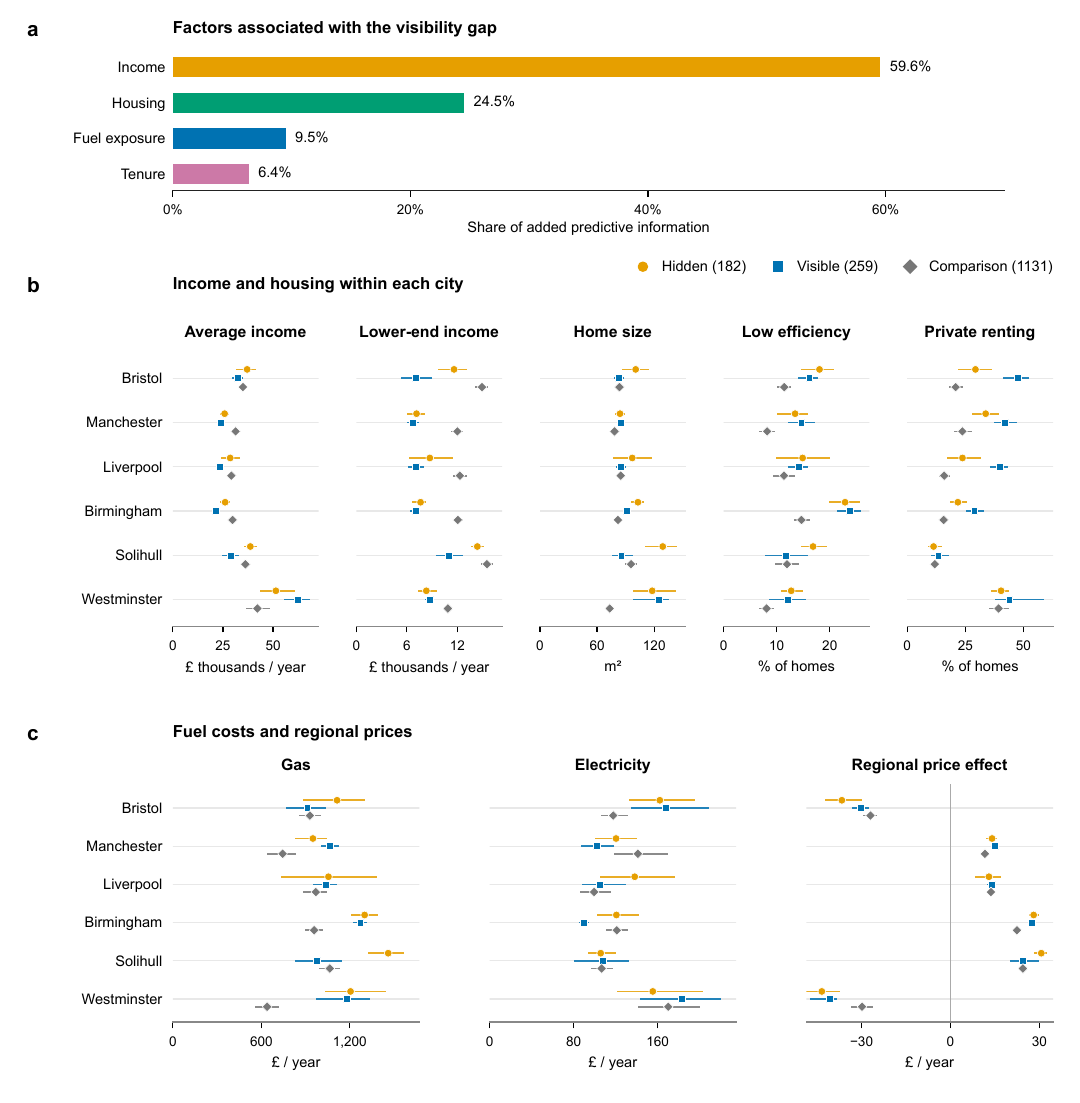}
\caption{Hidden exposure spans different income and housing configurations across cities. \textbf{a}, Share of additional cross-validated predictive information about the continuous exposure--visibility gap supplied by four variable groups, averaged over all entry orders after controls. \textbf{b}, City-specific profiles of hidden high-exposure areas (orange circles; $n=182$), visible high-exposure areas (blue squares; $n=259$) and the comparison group below both local upper-quarter cut-offs (grey diamonds; $n=1{,}131$). Average income and lower-end income are the fitted after-housing-cost mean and tenth percentile. Home size, EPC E--G share (low efficiency) and private renting are core-property summaries; renting uses the EPC tenure proxy. \textbf{c}, Gas and electricity structural-cost increments under common-reference Direct Debit prices, and the paired actual-region minus common-reference total increment (regional price effect). Increments are absolute annual structural-cost changes, averaged equally over S1--S3 relative to S0. Common-reference standing charges add approximately \pounds53 per year in every city. Panels \textbf{b,c} give equal weight to LSOAs within each city/group; points show means and lines 95\% intervals from 1,000 whole-MSOA bootstrap draws. Reference groups remain fixed. City-specific counts and supporting contrasts are reported in the Supplementary Information.}
\label{fig:associative-structure}
\end{figure}

The neighbourhoods identified by this comparison combined different income and housing conditions across cities (Fig.~\ref{fig:associative-structure}). A model evaluated on areas not used to fit it explained 77.8\% of variation in the exposure--visibility gap after accounting for data support, estimation method and local-authority differences. These adjustment variables alone explained 3.0\% of the variation in this gap. Income supplied 59.6\% of the additional predictive information and housing need and fabric 24.5\%; fuel transmission and tenure supplied smaller shares (Fig.~\ref{fig:associative-structure}a). These shares partition predictive information among correlated variable groups.

To place these income settings in national context, we compared neighbourhood means with an official benchmark. Of the 182 reference hidden-exposure neighbourhoods, 72 (39.6\%) had fitted mean equivalised disposable income after housing costs above \pounds34,500. Equivalisation adjusts income for household size and composition. This benchmark is the Office for National Statistics (ONS) median of mean incomes in Middle layer Super Output Areas (MSOAs) across England and Wales for the financial year ending (FYE) 2023 \cite{ONS2025SmallAreaIncomeBulletin}. Above-benchmark areas comprised 77.3\% of hidden neighbourhoods in Westminster, 76.9\% in Solihull and 60.5\% in Bristol, compared with 25.0\% in Liverpool, 12.8\% in Birmingham and 3.0\% in Manchester (Supplementary Table~S41).

The city profiles characterise the hidden geography by distinguishing average income from its lower end (Fig.~\ref{fig:associative-structure}b). Relative to the comparison group, hidden areas had lower fitted lower-end income in every city. Yet average income was higher in Westminster and Solihull. In Westminster, average income was approximately \pounds51,400 versus \pounds42,200, while lower-end income was \pounds8,300 versus \pounds10,900. Higher neighbourhood mean income therefore coexisted with exposure at the lower end of the fitted distribution. Manchester and Birmingham showed a different combination, with both average and lower-end incomes below those of the comparison group.

Housing differences further separated these configurations. Hidden areas had larger homes than the comparison group in Westminster (118 versus 73\,m$^2$) and Solihull (129 versus 96\,m$^2$); Manchester's contrast was smaller. Birmingham combined larger homes with more low-efficiency homes rated EPC E--G (22.9\% versus 14.7\%). Private renting was higher in hidden areas in Bristol, Manchester, Liverpool and Birmingham, with no clear contrast in Westminster or Solihull. Complete profiles also identify higher flat shares in Bristol's and Liverpool's hidden areas (Supplementary Information).

Changing the comparison to visible high-exposure areas revealed a different contrast. At comparable exposure, hidden areas had higher mean incomes and larger homes than visible high-exposure areas, but less private renting and no clear EPC E--G difference. The features distinguishing hidden exposure therefore depended on the comparison group (Supplementary Information, section 7).

To separate housing-related cost differences from regional tariffs, we applied the same reference prices to every area. Cost increases varied more across areas under these common prices than they changed when regional tariffs were applied (Fig.~\ref{fig:associative-structure}c). Under common-reference prices, annual gas-related increments in hidden areas ranged from approximately \pounds950 in Manchester to \pounds1,460 in Solihull. Electricity-related increments were higher in hidden areas than in the comparison group in Bristol and Liverpool, but lower in Manchester and Westminster. Actual regional tariffs shifted total increments in the hidden-exposure group by only \(-\)\pounds43 to \pounds31, reducing them in Bristol and Westminster. The common-price comparison revealed larger differences in housing-related cost increases than those introduced by regional tariffs.

\subsection{Fuel composition changes which neighbourhoods are hidden}

\begin{figure}[htbp]
\centering
\includegraphics[width=\linewidth]{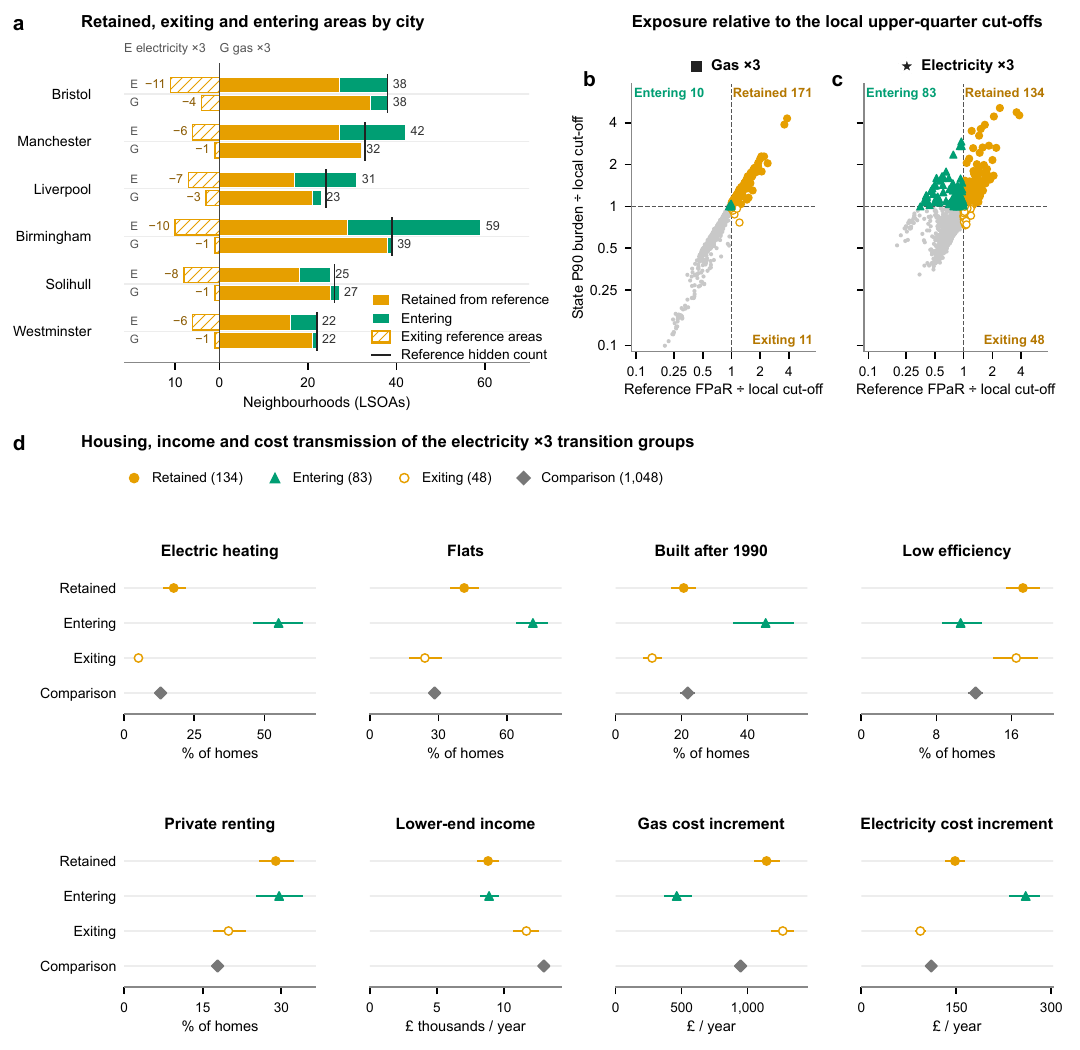}
\caption{Electricity-only stress changes which neighbourhoods rank as hidden high exposure. Entry and exit refer to local ranks. \textbf{a}, For each city, the reference hidden-exposure group (equal-weight S1--S3 FPaR; black tick) compared with the single-state hidden-exposure group under electricity-only (E) and gas-only (G) threefold unit prices, with the other unit price and standing charges at S0. Solid orange bars count retained reference areas, green bars count entering areas and hatched bars left of zero count exiting reference areas; labels give the single-state hidden count. \textbf{b,c}, Assessable LSOAs with official 2024 fuel poverty below the local top quarter, positioned by reference FPaR and single-state mean P90 burden. Each measure is divided by its within-authority upper-quarter cut-off (dashed lines at 1). Panels show gas-only (\textbf{b}) and electricity-only (\textbf{c}) threefold prices. Heading symbols match Fig.~\ref{fig:parametric-price-response}. Orange circles are retained, green triangles entering, open circles exiting and grey other assessable areas. \textbf{d}, Pooled equal-LSOA profiles of the four electricity-only groups: retained ($n=134$), entering ($n=83$), exiting ($n=48$) and the remaining comparison group below both cut-offs ($n=1{,}048$). Electric heating is the EPC main-fuel price carrier and includes direct and storage systems. Low efficiency is EPC E--G. Lower-end income is the fitted tenth percentile. Gas and electricity cost increments are common-reference structural-cost increments averaged over S1--S3 relative to S0, as in Fig.~\ref{fig:associative-structure}c. Points show means and lines 95\% intervals from 1,000 whole-MSOA bootstrap draws stratified by local authority. Reference FPaR groups remain fixed in Figs~\ref{fig:hidden-exposure-divergence}, \ref{fig:spatial-misalignment} and \ref{fig:associative-structure}.}
\label{fig:hidden-membership-transition}
\end{figure}

These housing and income profiles describe the crisis-based reference groups. We next tested whether changing the fuel-price balance changed which neighbourhoods belonged to the hidden-exposure group. Under 2026 tariffs, neighbourhood ordering remained close to reference FPaR ($\rho=0.994$); the hidden group again contained \CurrentHiddenCount{} areas, with Jaccard overlap \CurrentHiddenJaccard{} (Supplementary Table~S38). Tripling electricity prices while holding gas prices and standing charges at S0 broke this continuity (Fig.~\ref{fig:hidden-membership-transition}). Using each state's mean P90 burden, this stress identified \ElecHidden{} hidden areas. Entry and exit here concern local rank, not the onset or disappearance of hardship. Compared with the crisis-based group, \ElecRetained{} reference areas were retained, \ElecExiting{} exited and \ElecEntering{} entered (Fig.~\ref{fig:hidden-membership-transition}a,c). Entrants comprised 38.2\% of the electricity-led hidden group. Neighbourhood rank agreement with reference FPaR was $\rho=0.764$, and Jaccard overlap was 0.506. Gas-only threefold prices changed group membership much less, retaining \GasRetained{} areas, adding \GasEntering{} and removing \GasExiting{} from the high-exposure quarter (Jaccard 0.891; Fig.~\ref{fig:hidden-membership-transition}b). Every entering area came from the comparison group below both local cut-offs, and no exiting area remained in its local high-exposure quarter (Supplementary Table~S39). Birmingham supplied \ElecEnteringBirmingham{} of the \ElecEntering{} entering areas.

The entering and exiting areas had distinct housing configurations (Fig.~\ref{fig:hidden-membership-transition}d and Supplementary Table~S40). Electric heating served \EnterElectricShare\% of core properties in entering areas, compared with \RetainElectricShare\% in retained and \ExitElectricShare\% in exiting areas. Entering areas were dominated by flats (\EnterFlatShare\%), had the largest share of homes built after 1990 (\EnterNewShare\%) and the lowest EPC E--G share (\EnterEpcEG\%), with more social renting and less owner occupation than the retained group. Their gas-related cost increment under common prices was \pounds\EnterGasIncrement{} per year, against \pounds\RetainGasIncrement{} in retained areas, whereas their electricity-related increment was the largest of any group (\pounds\EnterElecIncrement{}). Fitted lower-end incomes in entering areas (\pounds\EnterLowerIncome) matched those of retained areas (\pounds\RetainLowerIncome). For context, the Department for Work and Pensions' FYE 2023 relative low-income benchmark after housing costs was approximately \pounds17,000 annually on a couple-equivalent basis \cite{DWP2025HBAIFye2023} (Supplementary Information, section 8). Exiting areas were instead gas-heated houses (\ExitFlatShare\% flats), mostly owner-occupied (\ExitOwnerShare\%), with lower-end incomes \pounds\ExitMinusRetainLowerIncome{} higher than in retained areas (95\% interval \pounds\ExitMinusRetainLowerIncomeLow--\ExitMinusRetainLowerIncomeHigh). Under electricity-only prices, their mean P90 burden reached \ExitMeanPninetyState\%, below their local high-exposure cut-off. The hidden-exposure list therefore depends on fuel-price composition: electricity-led prices brought newer, electrically heated rented flats into the hidden-exposure group while gas-heated houses with higher lower-end incomes left that high-exposure group. Responses of the fixed reference groups under the same prices are reported in Supplementary Fig.~S11c.

\clearpage

\section{Discussion}

Prospective assessment reveals both the extent of exposure beyond established hardship hotspots and changes in its housing profile. Across gas-led crisis reference prices, two in five locally high-exposure neighbourhoods lay outside the official fuel-poverty top quarter. Electricity-led stress then brought neighbourhoods with newer, more efficient housing, more flats and greater electricity dependence into the hidden-exposure group. These findings widen the scope of crisis preparedness to households outside current hotspots and housing configurations whose exposure changes with fuel prices.

\subsection{Hidden exposure beyond current hardship maps}

Official fuel poverty combines an efficiency threshold with a residual-income test under reference conditions \cite{DESNZ2026FuelPovertyMethodology,DESNZ2026SubregionalMethodology}. Prospective exposure instead traces required-energy burdens across local incomes and specified prices, covering all efficiency bands. This extends indicator-comparison research to prospective pressure \cite{RobinsonBouzarovskiLindley2018FuelPovertyIndicators,ChenRobinsonSingleton2025MultidimensionalEnergyDeprivation}; disagreement does not establish an official undercount. Moreover, 48.4\% of reference hidden areas already had high mean P90 burden at 2021 prices (Supplementary Table~S26). Hidden exposure includes intensification of existing modelled burdens as well as threshold crossing.

Hidden areas in Westminster and Solihull combined higher mean incomes with lower fitted lower-end incomes and larger homes than the comparison group; Birmingham combined lower incomes with larger homes and more inefficient stock (Fig.~\ref{fig:associative-structure}b). Earlier work linked larger absolute bill shocks in affluent areas to higher energy use \cite{FetzerGazzeBishop2024EnergyPriceShocks}. The income distributions here reveal how high burdens can arise within such areas. Comparisons with visible high-exposure areas also identify different tenure contrasts, consistent with interacting housing, income and tenancy conditions \cite{MiddlemissGillard2015EnergyVulnerability}.

\subsection{Spreading burdens and changing neighbourhood priorities}

Along the crisis price path, the neighbourhood count representing half of each authority's summed high-burden probabilities nearly doubled despite persistent property rankings (Table~\ref{tab:central-area-results}c). This connects persistent English fuel-poverty geographies with price-driven expansion beyond established Dutch hotspots \cite{BridgenRobinson2023DecadeFuelPoverty,MulderDallaLongaStraver2023SpatialEnergyPoverty}. July--September 2026 tariffs likewise left neighbourhood rankings closely aligned with the crisis reference while placing fewer properties at high burden than the unmitigated late-2022 cap.

Electricity-only threefold stress changed the housing profile. Entering neighbourhoods had newer, more efficient housing, more flats and greater electricity dependence than retained areas, despite similar lower-end incomes (Fig.~\ref{fig:hidden-membership-transition}). Fuel dependence altered priorities without requiring lower incomes or less efficient housing, a pattern that heating electrification can extend to more of the housing stock. Threshold sensitivity nevertheless depends on required costs relative to incomes as well as fuel shares (Fig.~\ref{fig:parametric-price-response}).

\subsection{Preparing support across fuel-price conditions}

Preparedness should include anticipatory outreach and contingency-support planning for households beyond current hardship hotspots. Fuel-specific stress tests can identify neighbourhoods for this additional screening and distinguish exposure across several price conditions from sensitivity to particular fuels. This extends attention to housing configurations, including efficient electrically heated stock, that established hardship geography may not prioritise. Local household evidence remains necessary to establish eligibility and support needs.

The appropriate response also depends on housing and tenure. Fabric improvement can reduce required energy; heating conversion changes both efficiency and fuel dependence. Tariff and income measures address affordability directly and warrant assessment for low-income electric flats. Occupancy and building condition inform retrofit assessment, while tenure identifies who pays costs and controls improvements. These uses complement municipal assessment frameworks \cite{PittalisEtAl2026UrbanEnergyPoverty}; delays in official small-area data reinforce the case for anticipatory screening \cite{DESNZ2026SubregionalFuelPoverty}.

\subsection{Interpretation, transferability and validation}

The baseline combines June 2026 EPC records with 2018 income-distribution shapes calibrated to FYE 2023 means. Required-energy costs retain standardised needs when households restrict consumption \cite{AndersonWhiteFinney2012ColdHomes,CongEtAl2022EnergyEquityGap}, but can exceed measured use \cite{SunikkaBlankGalvin2012Prebound,FewEtAl2023EPCOverPrediction}. Exposure depends on cost calibration and pairing dwellings with fitted local incomes under fixed demand. Occupants' actual incomes, transitions into fuel poverty and financial or behavioural resilience \cite{BurlinsonDavillasGiuliettiPrice2024PriceResilience} require household-level evidence. Income also enters the exposure measure, so predictive shares characterise the geography without establishing causal contributions.

Electrification requires updating this baseline. Existing direct and storage heating cannot stand in for heat pumps. An illustrative comparison holding delivered heat fixed, with a heat pump delivering 2.8 units of heat per unit of electricity over a season and boiler efficiency of 90\%, gives similar heating-cost increases from 2021 under the unmitigated late-2022 cap (11.9 versus 12.7 pence per kWh of heat), but a wider difference under July--September 2026 tariffs (2.5 versus 4.4 pence; Supplementary Information, section 9). Future applications should therefore recalculate energy requirements and fuel shares under retrofit and heating-replacement scenarios before retesting exposure.

Transfer requires dwelling-level energy needs by fuel, local income distributions, tariffs and a hardship benchmark. Income fitting extrapolates below the tenth and above the ninetieth percentiles and scales incomes uniformly within each MSOA. Spatial aggregation and classification cut-offs affected membership more than published mean-income uncertainty, favouring continuous exposure gaps over exact lists. Official rates are also modelled \cite{DESNZ2026SubregionalMethodology}, and uneven coverage, particularly Westminster's heat-network exclusions, restricts inference. Wider coverage, propagated income uncertainty, attribution of indicator disagreement and links to unmet needs, arrears and support demand would test practical usefulness.

These findings broaden the scope of crisis preparedness to include households outside current hardship hotspots and housing configurations whose exposure changes with fuel prices. Repeated assessments, updated for housing and heating changes, can guide where anticipatory outreach and contingency support may be needed.
\section{Methods}

\subsection{Study design and property sample}

We compared Bristol, Manchester, Liverpool, Birmingham, Solihull and Westminster using a theory-led design. Selection maximised contrasts in dwelling form, age, energy performance, floor area, tenure, heating fuel, density and deprivation. Westminster supplies a high-density case, Birmingham--Solihull and Manchester--Liverpool neighbouring-market contrasts, and Bristol an intermediate configuration.

The stress test uses today's assembled baseline, linking the latest-known EPC at 30 June 2026 to FYE 2023 calibrated income and official 2024 LILEE estimates \cite{DESNZ2026FuelPovertyMethodology}. Deprivation context uses the 2025 English Indices of Deprivation \cite{MHCLG2025EnglishIndicesDeprivation}.

The analytical unit pairs EPC property $i$, in LSOA $l$, MSOA $m$ and local authority district (LAD) $a$, with its fitted LSOA income distribution. EPC attributes supply the required-cost numerator and fitted income the denominator. Property counts denote EPC records, whereas exceedance probabilities describe the share of fitted local incomes at which required-energy burden exceeds the threshold. Official fuel poverty is an external small-area comparator.

We hold dwelling characteristics, required energy needs and fitted local income distributions fixed throughout the analysis. Gas and electricity unit prices vary independently, with standing-charge increments examined separately and complete tariffs evaluated as dated applications. State-specific analyses quantify burden, threshold entry and spatial coverage. The S1--S3 reference ensemble defines FPaR for comparison with official fuel poverty and for association and spatial analyses. Two further questions are kept distinct: how the fixed reference groups respond under other price states, and how single-state classification changes which neighbourhoods are identified.

Domestic Energy Performance Certificates (EPCs) were downloaded from the official bulk service and harmonised at 30 June 2026 \cite{MHCLG2026EPCData,MHCLG2025EPCTechnicalNotes,DLUHC2023EPCDataDictionary}. Across the six areas, 1,349,920 certificate events resolved to 1,064,208 latest properties through a composite normalised address-plus-postcode identity. The unique property reference number (UPRN) remained a secondary linkage and geography signal. We selected the latest certificate for each property and excluded properties whose selected certificate failed eligibility checks.

The coverage sample contains 1,033,255 properties across 1,781 LSOAs and 362 MSOAs. Eligibility, attrition and sample sensitivities appear in Supplementary Information, section 2; Table~\ref{tab:central-area-results} summarises the core sample, local income context and responses.

\subsection{Small-area income distributions}

ONS administrative statistics provide nine annual net occupied-address income quantiles, P10--P90, for tax year ending 2018 at 2011 LSOA level \cite{ONS2022AdminIncomeOccupiedAddresses}. These quantiles identify each LSOA's relative income position and shape. We fitted a finite-variance Dagum distribution to them using two-step feasible minimum quantile distance (MQD) \cite{Spasova2024MQD}. The official exact-fit lookup translates the fitted distributions to 2021 LSOAs, retaining mixtures where source areas were merged \cite{ONS2022LSOAExactFitLookup}.

We applied one common multiplicative scale factor to all fitted LSOA distributions within each MSOA. Their household-weighted mean then reproduced the official FYE 2023 MSOA mean equivalised disposable household income after housing costs \cite{ONS2025SmallAreaIncome}. This preserves within-MSOA differences in relative position and shape while calibrating the income level to 2023. Published upper and lower MSOA confidence limits provide shape-preserving income-anchor sensitivities.

All 1,744 source-area fits were usable, yielding calibrated income distributions for all 1,781 study LSOAs and 1,033,255 properties. Supplementary Information, section 3, reports the fitting objective, geographical harmonisation and fit diagnostics.

\subsection{Structural required-energy costs}

We used EPC ratings based on the Standard Assessment Procedure (SAP) to estimate structural required-energy costs under reference occupancy and service assumptions, consistent with the required-energy basis of official fuel-poverty assessment \cite{DESNZ2026FuelPovertyMethodology}. This basis retains standardised needs when observed consumption is affected by underheating or self-rationing. Inverting the rating yields a method-specific SAP-reference cost index expressed in GBP-year-equivalent units.

Let $\mathrm{EER}_i$ be the published integer energy efficiency rating, $A_i$ floor area and $v_i$ the inferred formula family. The EPC extract does not provide a complete, directly usable SAP or Reduced Data Standard Assessment Procedure (RdSAP) version field. We therefore infer the formula family from report type and lodgement date. RdSAP records switch from the 2012 family to RdSAP 10 on 15 June 2025; Full SAP records switch from SAP 2012 to SAP 10.2 on 15 June 2022. A sensitivity covers 7,271 transition-window Full SAP records. We invert the rating to obtain energy-cost factor $e_i$:
\begin{equation}
e_i=
\begin{cases}
(100-\mathrm{EER}_i)/13.95, & v_i\in\{\mathrm{SAP\ 2012,RdSAP\ 2012}\},\ \mathrm{EER}_i\geq52,\\
10^{(117-\mathrm{EER}_i)/121}, & v_i\in\{\mathrm{SAP\ 2012,RdSAP\ 2012}\},\ \mathrm{EER}_i<52,\\
(100-\mathrm{EER}_i)/16.21, & v_i\in\{\mathrm{SAP\ 10.2,RdSAP\ 10}\},\ \mathrm{EER}_i\geq44,\\
10^{(108.8-\mathrm{EER}_i)/120.5}, & v_i\in\{\mathrm{SAP\ 10.2,RdSAP\ 10}\},\ \mathrm{EER}_i<44.
\end{cases}
\end{equation}
The structural baseline is
\begin{equation}
C_{i0}=\frac{e_i(A_i+45)}{\kappa_{v_i}},
\end{equation}
where $\kappa_{v_i}=0.42$ for SAP/RdSAP 2012 and RdSAP 10 and $\kappa_{v_i}=0.36$ for SAP 10.2. The SAP/RdSAP 2012 parameters follow the official 2012 procedure \cite{BRE2014SAP2012}; the SAP 10.2 parameters follow its specification \cite{BRE2025SAP10}; and the RdSAP 10 assignment uses the SAP 10 rating equation with the RdSAP deflator documented in the RdSAP specification \cite{BRE2026RdSAP10}. The integer rating is used directly. All 23 retained score-100 properties consequently have $C_{i0}=0$ and remain visible, with a prespecified exclusion sensitivity.

EPC heating, hot-water and lighting costs determine the fuel shares used to reprice the reconstructed structural baseline. If $H_{ij}$ is service $j$ cost and $j$ maps to fuel $f$,
\begin{equation}
w_{if}=\frac{\sum_jH_{ij}\mathbf{1}(j\mapsto f)}{\sum_jH_{ij}}.
\end{equation}
Positive-cost services enter the denominator, preserving the supported share. Repricing covers separately costed gas, electricity or oil pathways.

Service mapping and price support yield 962,869 core gas/electricity properties, 723 oil properties and 69,663 partial or unsupported properties. FPaR-based classification uses the core gas/electricity sample. Supplementary Information, section 5, reports method, service and eligibility diagnostics.

\subsection{Tariff inputs and price-space design}

Ofgem workbooks supply regional gas and single-register electricity cap components \cite{Ofgem2026DefaultTariffCapLevels}; official tables supply late-2022 Energy Price Guarantee rates \cite{DESNZ2022EPGRegionalRates}. Postcode-assigned regional Direct Debit prices are primary; Standard Credit and Prepayment provide sensitivities.

We assigned Ofgem regions by overlaying May 2026 ONS postcode points on distribution-network licence areas published by the National Energy System Operator (NESO) \cite{ONS2026PostcodeDirectory,NESO2024DNOLicenceAreas}. Grid Supply Point regions supplied an independent check \cite{NESO2025GSPRegions}. For each fuel, payment method and period, common-reference unit rates and standing charges are arithmetic means of six local-authority reference-region contributions. Birmingham and Solihull each contribute one-sixth through Midlands. Applying common prices to the same properties separates housing-related cost increases from regional tariff differences.

For fuel $f$, state $s$, price geography $g$ and payment method $q$, unit prices are $p_{fsgq}$ and daily standing charges for regulated meter $k$ are $SC_{ksgq}$.

\begin{samepage}
Price ratios and structural-cost changes are expressed relative to matching S0 components:
\begin{align}
r_{fsgq}&=\frac{p_{fsgq}}{p_{f0gq}},\\
\Delta U_{isgq}&=C_{i0}\sum_f w_{if}(r_{fsgq}-1),\\
\Delta S_{isgq}&=\frac{365}{100}\sum_{k\in\mathcal{M}_i}(SC_{ksgq}-SC_{k0gq}),\\
C_{isgq}&=\max\{0,C_{i0}+\Delta U_{isgq}+\Delta S_{isgq}\}.
\end{align}
\end{samepage}
Every eligible dwelling includes an electricity meter because fixed lighting uses electricity; gas enters where the main carrier or a positive-cost service uses mains gas. Repricing adds changes in unit and standing charges relative to structural S0.

\begin{samepage}
The joint price design varies gas and electricity unit prices independently, with standing-charge increments examined in separate slices. For gas multiplier $g$, electricity multiplier $e$ and an annual increment $a$ applied to each active regulated meter,
\begin{equation}
C_i(g,e,a)=C_{i0}\!\left[1+w_{i,\mathrm{gas}}(g-1)+w_{i,\mathrm{electricity}}(e-1)\right]
+a\!\sum_{k\in\{\mathrm{electricity,gas}\}}\mathbf{1}(k\in\mathcal{M}_i).
\end{equation}
\end{samepage}
The primary surface evaluates $g=1.00,1.25,\ldots,4.50$ and $e=1.00,1.20,\ldots,3.00$ at $a=0$. These bounds contain every area-specific S0--S3 unit-rate coordinate, including the approximately 2.7-fold electricity increase under the common-reference crisis cap. Holding the gas price fixed isolates an electricity-only stress. The zero-increment surface holds standing charges at S0; complete tariffs apply each state's unit rates and standing-charge changes.

Dated tariffs locate applications within this space. S0 covers April--September 2021; S1, October 2021--March 2022; and S2, April--September 2022. The wholesale-price observation window informing S2 closed in January 2022, before the escalation of the Russia--Ukraine conflict in February 2022 \cite{Ofgem2022WholesaleMethodology}. S3 is the unmitigated October--December 2022 underlying-cap counterfactual. S4 combines historical Energy Price Guarantee unit rates with S3 standing charges. S5 denotes the separate heating-oil stress reported in Supplementary Table~S22. S6a--S6c cover January--March, April--June and July--September 2026, respectively. Unit-rate coordinates are ratios to each area's matching S0 prices.

The fifteen-state comparison combines eight complete gas/electricity tariffs with seven additional stresses (Supplementary Table~S37). The additional states comprise gas-only, electricity-only and common threefold S0 unit prices, a fixed-charge stress, and three further S6c stresses. The latter raise gas, electricity or both unit prices by 50\%, retaining S6c standing charges. These comparisons use the same core properties and fitted incomes under actual-region Direct Debit pricing.

Supplementary Information, sections 4 and 6, report tariff extraction, historical context, heating-oil comparisons, marginal price experiments, fixed-charge sensitivities and proportional scaling.

\subsection{Burden, threshold entry and spatial coverage}

\label{sec:methods-burden}

Let $Y_l$ follow the fitted LSOA income distribution, with cumulative distribution $F_l$ and quantile function $Q_l$. We set the analytical high-burden threshold to $\tau=0.10$, following the historical expenditure-to-income benchmark in fuel poverty research \cite{Boardman1991FuelPoverty,Moore2012DefinitionsFuelPoverty}. For a fixed property cost, the 90th-percentile required-energy burden equals that cost divided by the fitted local income tenth percentile. We refer to this upper-tail quantity as P90 burden and use it to assess affordability towards the lower-income end. The share of properties with P90 burden at or above 10\% measures how widely this exposure extends across the housing stock. Baseline P90 burden is
\begin{equation}
B^{0}_{i,.90}=\frac{C_{i0}}{Q_l(0.10)}.
\end{equation}

For each historical, counterfactual, contemporary or plausible stress state $s$, P90 burden and fitted-income threshold probability are
\begin{align}
B_{is,.90}&=\frac{C_{is}}{Q_l(0.10)},\\
P_{is}(B>\tau)&=F_l\!\left(\frac{C_{is}}{\tau}\right).
\end{align}
The fifteen-state comparison reports property and LSOA mean P90 burden, threshold probabilities and spatial coverage. Price-space summaries report each LAD, the property-weighted six-area result and an equal-LAD comparator.

The first-entry path uses the ordered baseline-to-final-stress states because P90 burden is non-decreasing along that sequence. When the threshold is reached, the entry state is
\begin{equation}
s_i^{\ast}=\min\left\{s\in\{\mathrm{S0,S1,S2,S3}\}:B_{is,.90}\geq\tau\right\}.
\end{equation}
Properties below the threshold through S3 form a separate class. S4 and S6a--S6c are separate complete-tariff comparisons outside the monotone path. Contemporary results apply updated tariffs to the same property and fitted-income baseline.

Baseline groups are fixed before stress. P90 burden bands are $<5\%$, 5--$<10\%$, 10--$<20\%$ and $\geq20\%$. Five near-equal baseline-burden fifths, labelled Q1--Q5 from lowest to highest, are assigned by sorting first on burden and then on a pseudonymous property key to resolve ties. Their approximate boundaries are 3.93\%, 5.29\%, 6.83\% and 9.57\%. We report movement across bands and groups, rank correlations, and retention in the highest fifth and tenth. Group responses are also reported by local authority, official fuel-poverty band, fuel, EPC band, property type and tenure proxy; official fuel-poverty groups use the 1,754 assessable areas.

To measure spatial reach, we use scenario-specific high-burden probabilities, which can be summed across properties. For LSOA $l$ with $n_l$ core properties, the high-burden mass $M_{ls}$ and mean probability $\bar P_{ls}$ are
\begin{equation}
M_{ls}=\sum_{i\in l}P_{is}(B>\tau), \qquad \bar P_{ls}=\frac{M_{ls}}{n_l}.
\end{equation}
Within each LAD $a$ and state $s$, LSOAs $\mathcal{L}_a$ are ordered as $l_{(1)},\ldots,l_{(|\mathcal{L}_a|)}$ by decreasing $\bar P_{ls}$. LSOA codes resolve ties.

For target coverage $c\in(0,1]$, we calculate
\begin{align}
k_{as}(c)&=\min\left\{k:\sum_{j=1}^{k}M_{l_{(j)}s}\geq c\sum_{l\in\mathcal{L}_a}M_{ls}\right\},\\
K_s(c)&=\sum_a k_{as}(c), \qquad \theta_{as}(c)=\bar P_{l_{(k_{as}(c))}s}.
\end{align}
Here $k_{as}(c)$ is the number of neighbourhoods selected, $K_s(c)$ the six-area count and $\theta_{as}(c)$ the boundary probability. Targets are applied within each authority before summing counts; coverage is the cumulative share of modelled exceedance probability mass. Supplementary Information, section 6, reports the grid, sensitivities and group composition.

To express structural threshold proximity at small-area scale, let
\begin{align}
\bar B_l^0&=\operatorname{mean}_{i\in l}\!\left(\frac{C_{i0}}{Q_l(0.10)}\right),\\
\bar G_l&=\operatorname{mean}_{i\in l}\!\left(\frac{C_{i0}}{Q_l(0.10)}w_{i,\mathrm{gas}}\right),\\
\bar E_l&=\operatorname{mean}_{i\in l}\!\left(\frac{C_{i0}}{Q_l(0.10)}w_{i,\mathrm{electricity}}\right).
\end{align}
Holding the other unit price at S0 and $a=0$, the critical gas, electricity and common multipliers are
\begin{equation}
g_l^\ast=1+\frac{\tau-\bar B_l^0}{\bar G_l},\qquad
e_l^\ast=1+\frac{\tau-\bar B_l^0}{\bar E_l},\qquad
m_l^\ast=\frac{\tau}{\bar B_l^0}.
\end{equation}
A missing fuel-specific value denotes zero exposure to that carrier; values at or below one denote mean LSOA burden already at or above 10\% under S0. These multipliers and the property-level headroom retained in the Supplementary Information are fixed-demand, fixed-income structural thresholds.

\subsection{Reference FPaR and exposure--visibility gap}

\label{sec:methods-reference-fpar}

Fuel Price-at-Risk (FPaR) summarises upper-tail burden jointly over fitted local incomes and the S1--S3 reference ensemble. The state index $S$ takes S1, S2 and S3 with equal analytical weights of one third. These weights specify the stress-test design and do not represent probabilities of future price states.

The reference FPaR is
\begin{equation}
\operatorname{FPaR}_{i,.90}=Q_{.90}\!\left(\frac{C_{iS}}{Y_l}\right),
\end{equation}
where the quantile is taken over the equally weighted mixture of state-specific burden distributions, each evaluated over fitted local income. We calculate FPaR from this mixture rather than averaging the state-specific P90 burdens.

For each LSOA, $R_l$ is the unweighted mean of reference FPaR across its core properties. Local-authority summaries weight property metrics by core-property count. This retains the property as the exposure unit while keeping local data support explicit.

Core support in LSOA $l$ is $\rho_l=n_l^{\mathrm{core}}/n_l^{\mathrm{coverage}}$. LSOAs with $\rho_l\geq0.80$ form the high-support stratum; those with $0.50\leq\rho_l<0.80$ form the bounded-support stratum; and those below 0.50 are not assessable for hidden exposure. The partition contains 1,663, 91 and 27 LSOAs, respectively.

Let $f_l$ be the official 2024 fuel-poverty rate. The official estimates provide the external small-area comparator and are regression-modelled for broad spatial comparison \cite{DESNZ2026SubregionalFuelPoverty,DESNZ2026SubregionalMethodology}. We compare the two measures within local authority $a$ using robust scales. The continuous exposure--visibility gap is
\begin{align}
z^{R}_l&=\frac{\ln(1+R_l)-\operatorname{median}_a\{\ln(1+R)\}}{\operatorname{IQR}_a\{\ln(1+R)\}},\\
z^{f}_l&=\frac{f_l-\operatorname{median}_a(f)}{\operatorname{IQR}_a(f)},\\
D_l&=z^{R}_l-z^{f}_l.
\end{align}
Positive $D_l$ indicates that price-contingent exposure is high relative to official visibility in the same local authority.

The reference hidden-exposure group is defined by
\begin{equation}
H_l=\mathbf{1}\!\left\{R_l\geq Q_{a,.75}(R),\ f_l<Q_{a,.75}(f)\right\}.
\end{equation}
This group contains LSOAs in the highest within-LAD FPaR quarter whose official fuel-poverty rates fall below the local upper-quartile threshold. A narrower restriction, $R_l\geq Q_{a,.75}(R)$ and $f_l\leq Q_{a,.25}(f)$, is retained in the Supplementary Information as a sensitivity. Both provide LSOA-level local-relative classifications.

We report both mirror groups: high-exposure LSOAs outside the fuel-poverty top quarter, and top-quarter fuel-poverty LSOAs outside the high-exposure quarter. The matched upper-quarter partitions give equal counts. Agreement is benchmarked against independence. For LAD $a$ with $n_a$ assessable LSOAs, of which $r_a$ are high-exposure and $h_a$ high-fuel-poverty, the coincidence expected under independence is $r_ah_a/n_a$, summed across LADs. Cohen's $\kappa$ summarises agreement beyond chance within each LAD, and the pooled value uses within-LAD expected agreement.

To trace high-burden reach, let $\bar B_{ls,.90}$ be mean property P90 burden in LSOA $l$ under state $s$. The reference groups comprise 182 hidden areas ($H_l=1$) and 259 visible high-exposure areas ($R_l\geq Q_{a,.75}(R)$ and $f_l\geq Q_{a,.75}(f)$). The comparison group comprises 1,131 areas below both local upper-quarter cut-offs ($R_l<Q_{a,.75}(R)$ and $f_l<Q_{a,.75}(f)$). Within each fixed group, we trace mean burdens across the fifteen states. Reach is the share of LSOAs with $\bar B_{ls,.90}\geq\tau$; the fixed-group paths are reported in Supplementary Table~S26 and Supplementary Fig.~S11c. The 1,131 areas below both cut-offs are labelled Comparison in Figure 5, and their non-entering subset in Figure 6d. Other assessable areas in Figures 3 and 4 additionally include the 182 high-official, lower-exposure mirror areas.

For each of the fifteen states, we compare property P90-burden ranks with S0 and assessable-LSOA mean-P90 ranks with S0 and reference FPaR. We then reapply the within-LAD upper-quartile rule and official-fuel-poverty boundary to scenario-specific mean P90 burden. We report high-burden and hidden-exposure group counts and Jaccard overlap with S0 and reference FPaR.

To show how hidden-exposure group membership changes with fuel composition, we compared each state's single-state group with the fixed reference group. Retained areas belong to both, entering areas only to the single-state group and exiting areas only to the reference group. Because the official fuel-poverty boundary is unchanged, entry and exit depend only on whether an area's mean P90 burden crosses its within-LAD upper-quarter cut-off; we display this as burden divided by the local cut-off. For the electricity-only and gas-only threefold conditions, we profiled the retained, entering, exiting and remaining comparison groups using the housing, income and common-price cost variables described below, with equal LSOA weight pooled across the six authorities and 1,000 whole-MSOA bootstrap draws stratified by local authority.

\subsection{Association and spatial analyses}

\label{sec:methods-associations}

We modelled predictive associations between $D_l$ and four prespecified variable groups: lower-income tail; housing need and fabric; fuel-price transmission; and tenure or limited-control proxies. The primary continuous association was ordinary least squares across all 1,754 assessable LSOAs, with local-authority fixed effects and standard errors clustered over 362 MSOAs. Core support and estimation-method composition were always included. Five-fold cross-validation kept MSOAs intact and stratified folds by local authority \cite{RobertsEtAl2017CrossValidation}. Each block's contribution was its mean sequential increase in out-of-sample $R^2$ across all $4!=24$ entry orders \cite{Gromping2007RelativeImportance}. A clustered logistic model of membership in the broad hidden-exposure group provided a secondary binary diagnostic; the strict class was not fitted because it did not meet the prespecified minimum class count.

City profiles distinguish hidden high-exposure areas, visible high-exposure areas and a comparison group below both local upper-quartile thresholds. Within each city and group, we average LSOA summaries of fitted mean income, lower-end income (the fitted tenth percentile), mean core-property floor area, EPC E--G share and private-rent-proxy share. All profiles retain their original units and give equal weight to LSOAs. Comparing hidden areas with the comparison group describes exposure within the same official-visibility band; comparing them with visible high-exposure areas describes differences among locally high-risk areas. We report their exposure distributions alongside a supplementary pooled comparison conditional on local-authority effects and city-specific slopes in log reference FPaR.

We separate fuel-price transmission from income by expressing gas-unit, electricity-unit and standing-charge increments in annual structural-cost units. For each core property, the common-reference calculation applies the same six-LAD reference prices to its unchanged structural baseline, service weights and meter configuration, then averages its S1--S3 increments with equal design weights. We aggregate first across core properties within each LSOA and then equally across LSOAs in each city and fixed group. Regional-price contributions are the paired difference between actual-region and common-reference increments and are reported separately. Gas, electricity and standing-charge contributions, including the zero-floor adjustment, sum to the total annual structural-cost increment. FPaR is evaluated separately as an upper-tail quantile.

Intervals for group means and contrasts use 1,000 whole-MSOA bootstrap draws within each local authority, holding fitted income distributions and class labels fixed. Intervals use draws containing all relevant groups and describe spatial sampling variation conditional on the model. Supplementary Information, section 7, reports the complete standardised profiles, fixed-label sensitivities and reference FPaR summaries by dwelling type, EPC band, tenure proxy and fuel (Supplementary Fig.~S4c).

Spatial structure in the continuous gap $D_l$ is tested separately within each local authority using queen-contiguity weights, which treat areas sharing an edge or corner as neighbours, on official ONS 2021 LSOA boundaries \cite{ONS2021LSOABoundaries}. Moran's $I$ uses 999 permutations \cite{Moran1950StochasticPhenomena}; two-sided permutation $p$-values are adjusted across the six tests using the Benjamini--Hochberg false-discovery-rate procedure \cite{BenjaminiHochberg1995FDR}. Each assessable geometry has at least one neighbour.

\subsection{Sensitivity analyses}

One-factor sensitivities vary the MSOA income anchor, equal-LAD pooled classification, price geography, payment method, EPC validity, strict-UPRN inclusion and SAP/RdSAP method assignment. Additional tests vary minimum core-support thresholds and the P70/P80 risk and official-visibility boundaries. Identity overlap is reported with the Jaccard index for the reference hidden-exposure group and, separately, for the narrower P25 official-visibility sensitivity.

We tested ten S1--S3 weighting schemes: equal weights, three schemes assigning all weight to one state, three dividing weight equally between two states, and three retaining all states with unequal weights. Each of the latter three schemes assigns 0.8 weight to one state and 0.1 to each other state. For every scheme, we recomputed property FPaR, aggregated it to LSOA means and reapplied the unchanged support-aware within-LAD comparison with official fuel poverty. We compared LSOA ranks, highest-quarter membership and broad and narrow hidden-exposure membership with the equal-weight centre using Spearman correlation and Jaccard overlap.

Profile sensitivities apply high-support and income-fit-boundary exclusions while retaining the original class labels. Full robustness results are reported in Supplementary Information, section 8.

\section*{Code and data availability}
\label{sec:code-data-availability}

Code and data are available at \url{https://github.com/DanialBai/FPaR}. The repository provides a bundle containing analysis code, processed aggregate data and figure Source Data for the main and supplementary figures and numerical tables, with software dependencies, data definitions and file checksums.

\section*{Competing Interests}

The authors declare no competing interests.

\section*{Acknowledgements}

The authors would like to thank the funding support from Westminster City Council and King's College London.

\section*{Author Contributions}

Yunfei Bai: Writing -- original draft, Conceptualization, Visualization, Methodology, Formal analysis, Data curation.
Aaron Tesfa Tsion: Writing -- review \& editing, Validation.
Ra\'ul Rosales and Barbara Shollock: Writing -- review \& editing, Project administration, Funding acquisition.
Wei He: Writing -- review \& editing, Conceptualization, Supervision, Project administration, Funding acquisition, Resources.

\bibliographystyle{naturemag}
\bibliography{shared/references}

\end{document}